\documentclass[a4paper]{jpconf}
\usepackage{graphicx}
\usepackage{float}
\begin{document}

\title{Exploring Replica-Exchange Wang-Landau sampling in higher-dimensional parameter space}

\author{Alexandra Valentim$^{1,2}$, Julio C. S. Rocha$^{1,3}$, Shan-Ho Tsai$^{1,4}$, Ying Wai Li$^5$, Markus Eisenbach$^5$, Carlos E. Fiore$^{2,6}$, David P. Landau$^1$}

\address{$1$ Center for Simulational Physics, University of Georgia, Athens, GA 30602, USA} 

\address{$2$ Departamento de F\'{\i}sica, Universidade Federal do Paran\'a, Curitiba-PR  81531-980, Brazil} 

\address{$3$ Departamento de F\'{\i}sica, Universidade Federal de Minas Gerais, Belo Horizonte-MG 31270-901, Brazil} 

\address{$4$ Enterprise Information Technology Services, University of Georgia, Athens, GA 30602, USA}

\address{$5$ National Center for Computational Sciences, Oak Ridge National Laboratory, Oak Ridge, TN 37831, USA} 

\address{$6$ Instituto de F\'{\i}sica, Universidade de S\~{a}o Paulo, S\~{a}o Paulo-SP 05315-970, Brazil}

\ead{alexandra@fisica.ufpr.br}

\begin{abstract}
We considered a higher-dimensional extension for the replica-exchange Wang-Landau algorithm to perform a random walk in the energy and magnetization space of the two-dimensional Ising model.
This hybrid scheme combines the advantages of Wang-Landau and Replica-Exchange algorithms, and the one-dimensional version of this approach has been shown to be very efficient and to scale well, up to several thousands of computing cores. This approach allows us to split the parameter space of the system to be simulated into several pieces and still perform a random walk over the entire parameter range, ensuring the ergodicity of the simulation.
Previous work, in which a similar scheme of parallel simulation was implemented without using replica exchange and with a different way to combine the result from the pieces, led to discontinuities in the final density of states over the entire range of parameters. From our simulations, it appears that the replica-exchange Wang-Landau algorithm is able to overcome this difficulty, allowing exploration of higher parameter phase space by keeping track of the joint density of states.
\end{abstract}

\section{Introduction}
The Wang-Landau (WL) Monte Carlo method \cite{WLprl} is a sampling technique that can be used to obtain the density of states (DOS) of a system. It has the property of generating a flat histogram in some random walk space, where the parameters for the random walk and the flatness criterion can be chosen according to the system of interest \cite{WLpre,WLajp}. Due to these characteristics and to the simplicity of the algorithm, the Wang-Landau approach has been studied and applied successfully to different kinds of systems (see, for examples, \cite{rathore,yamaguchi,taylor,wuest}), including those with rough free energy landscapes and strong first-order phase transitions, which cannot be easily studied with traditional methods. 

Recently an efficient parallel scheme of the Wang-Landau method was implemented \cite{REWLprl,REWLpre,REWLjop,TYVL}, whereby independent random walks are carried out in different overlapping windows of energy and a replica-exchange scheme \cite{marinari,nemoto} is used to exchange system configurations of neighboring energy windows. This parallel replica exchange Wang-Landau (REWL) method was shown to scale extremely well up to several thousands of computing cores. Therefore, it can be used to study more complex systems and much larger system sizes than it is possible with the serial Wang-Landau method. 

While the density of states as a function of a single parameter, such as the total energy, suffices to provide insight into many systems, there are a number of physical properties and systems whose investigation relies on a joint density of states dependent on additional parameters \cite{zhou,asymising,pla}. Therefore, it would be of great interest to extend the one-dimensional REWL method \cite{REWLprl,REWLpre,REWLjop} to a higher-dimensional parameter space, and implement a similar method that combines 
enhanced sampling techniques \cite{marinari,nemoto,REMUCAandMUCARE} and the advantages of  parallel computing to obtain a joint density of states.

In this exploratory study we consider the Ising model, for which the density of states as a function of total energy is known exactly \cite{beale}.
The phase space of this model, as a function of energy and magnetization, has cusps and empty regions with no realizable configurations.
Simulations in this higher-dimensional parameter space are challenging because, depending on the sampling method used,
the lower and higher energy edges of the phase space are hard to sample and thus have poor statistics. Moreover, the empty region between cusps can be hard to overcome in a simulation.

\section{Two-dimensional Replica-Exchange Wang-Landau sampling}
The REWL method \cite{REWLprl} is arguably the most successful parallelization attempt to the WL method so far, and it consists of only three simple modifications to the serial version of the algorithm.
Taking the Ising model as an example, a serial WL sampling of the total energy ($E$) and magnetization ($M$) phase space employs a two-dimensional random walk in the entire $E$ and $M$ ranges. In the process a random walker keeps track of the joint density of states $g(E,M)$ and the histogram of visited states $H(E,M)$.
Let $A'(E',M')$ and $A''(E'',M'')$ denote, respectively, the current and trial spin configurations of the system.
The WL probability that the trial move is accepted is given by
\begin{equation}
  P(A'\rightarrow A'') = \min \left[ 1, \,\frac{g(E',M')}{g(E'',M'')}\right].
\label{p-wl}
\end{equation}
If $A''(E'',M'')$ is accepted, the density of states $g(E'',M'')$ is adjusted by the modification factor $f>1$, that is, $g(E'',M'')\rightarrow g(E'',M'')*f$; and the histogram is updated with $H(E'',M'')\rightarrow H(E'',M'')+1$. If the trial state is not accepted, we update $g(E',M')\rightarrow g(E',M')*f$ and  $H(E',M')\rightarrow H(E',M')+1$. One Monte Carlo step is defined as $L\times L$ trial state considerations.

The random walker proceeds with choosing a new trial spin configuration and checking whether it is accepted, until $H(E,M)$ is considered flat, in accordance with a chosen criterion. At this point the current iteration is ended and the histogram is reset to zero.
The next iteration begins with a finer $f$, reduced by some predefined rule (e.g. $ f\rightarrow {\sqrt f}$ ). More iterations are performed until the modification factor reaches a chosen final value $f_{\rm final}$, such as $f_{\rm final}=\exp(10^{-8})$. 
The density of states is never reset and, at the end of the simulation, $g(E,M)$ is expected to converge to the true value with an accuracy proportional to $\sqrt{ f_{\rm final}}$.

\begin{figure}[h]
\centering
\begin{minipage}{18pc}
\includegraphics[width=18pc]{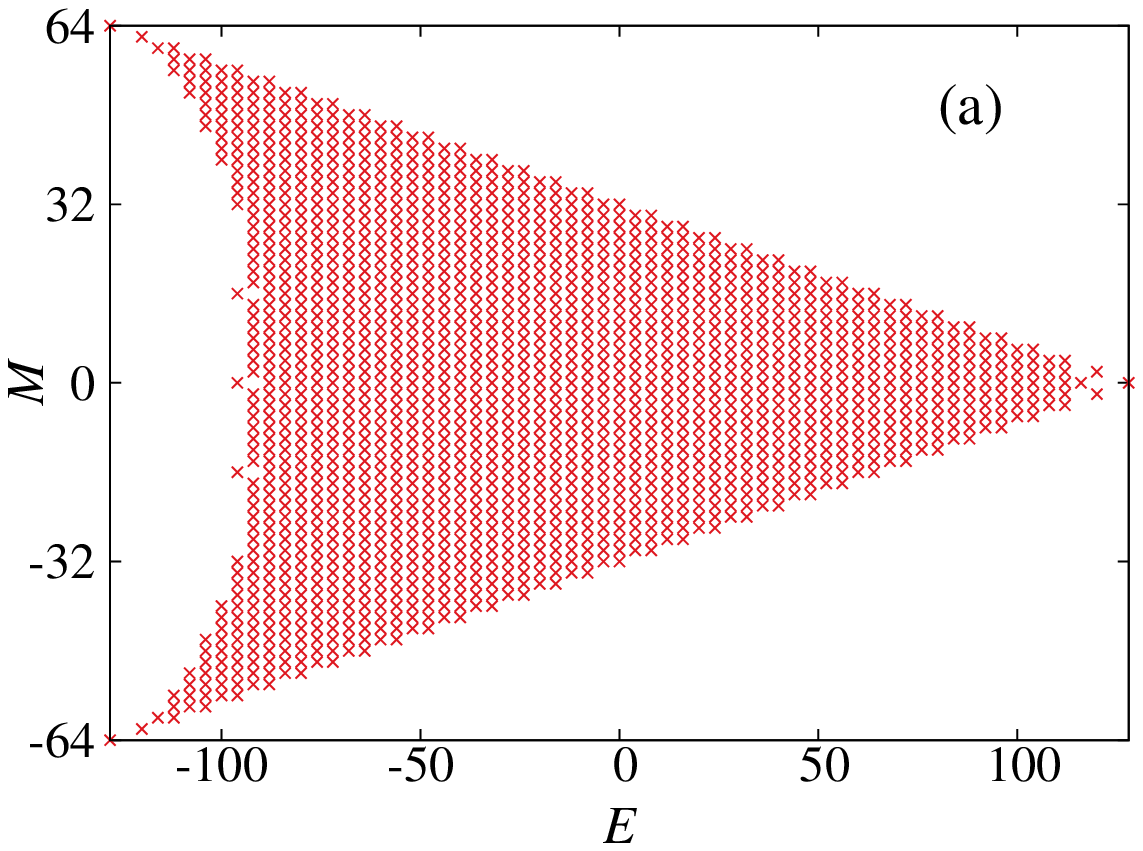}
\end{minipage}\hspace{0pc}%
\begin{minipage}{18pc}
\includegraphics[width=22pc]{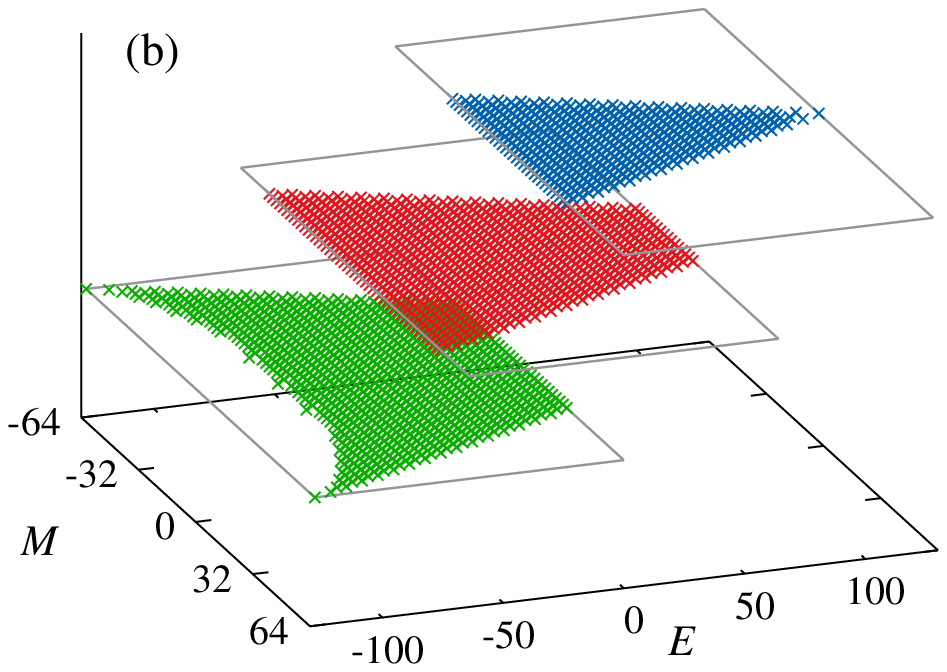}
\end{minipage} 
\caption{(color online) (a)  Complete phase space for the Ising model on a $8\times 8$  square lattice. (b) An example of splitting of the parameter space into overlapped, equal-sized energy windows. Energy windows are shifted upward progressively for clarity.}
\label{fig1}
\end{figure}

A replica-exchange Wang-Landau sampling starts by splitting the phase space into smaller overlapping subspaces (windows), as schematically shown in Fig. \ref{fig1}.
For simplicity, only the energy range is split and the subspaces were artificially displaced vertically (in Fig.\ref{fig1}(b)) for clarity. 

Inside each subspace one or more independent random walkers perform a standard WL sampling, with independent $g(E,M)$ and $H(E,M)$.
After a predetermined number of Monte Carlo steps, a configurational exchange between two walkers in adjacent windows is proposed. The walker pairs are chosen at random. If either of them is not in the overlapping region, the exchange of configurations will not be attempted. However, if the walker pairs are in the same overlapping region, the configuration swap between them (walkers $i$ and $j$) will be accepted with the probability:
\begin{equation}
  P_{\rm acc}= \min \left[ 1, \,\frac{g_i(E(X),M(X))}{g_i(E(Y),M(Y))}\frac{g_j(E(Y),M(Y))}{g_j(E(X),M(X))}\right],
\label{p-acc}
\end{equation}
where $X$ and $Y$ denote, respectively, the spin configurations of walkers $i$ and $j$ before the exchange. Note that $P_{\rm acc}$ satisfies detailed balance. All walkers (in all regions) move to the next iteration at the same time and the parallel simulation ends when the modification factor for every walker reaches $f_{\rm final}$. 

\begin{figure}[h]
\centering
\begin{minipage}{19pc}
\includegraphics[width=20pc]{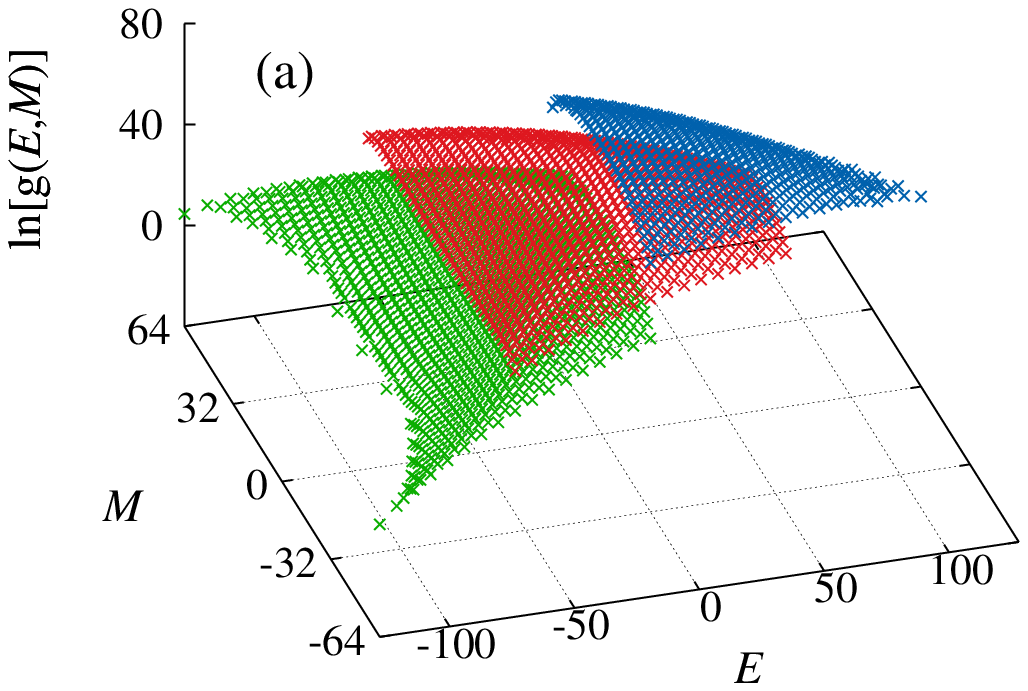}
\end{minipage}\hspace{0pc}
\begin{minipage}{18pc}
\includegraphics[width=18pc]{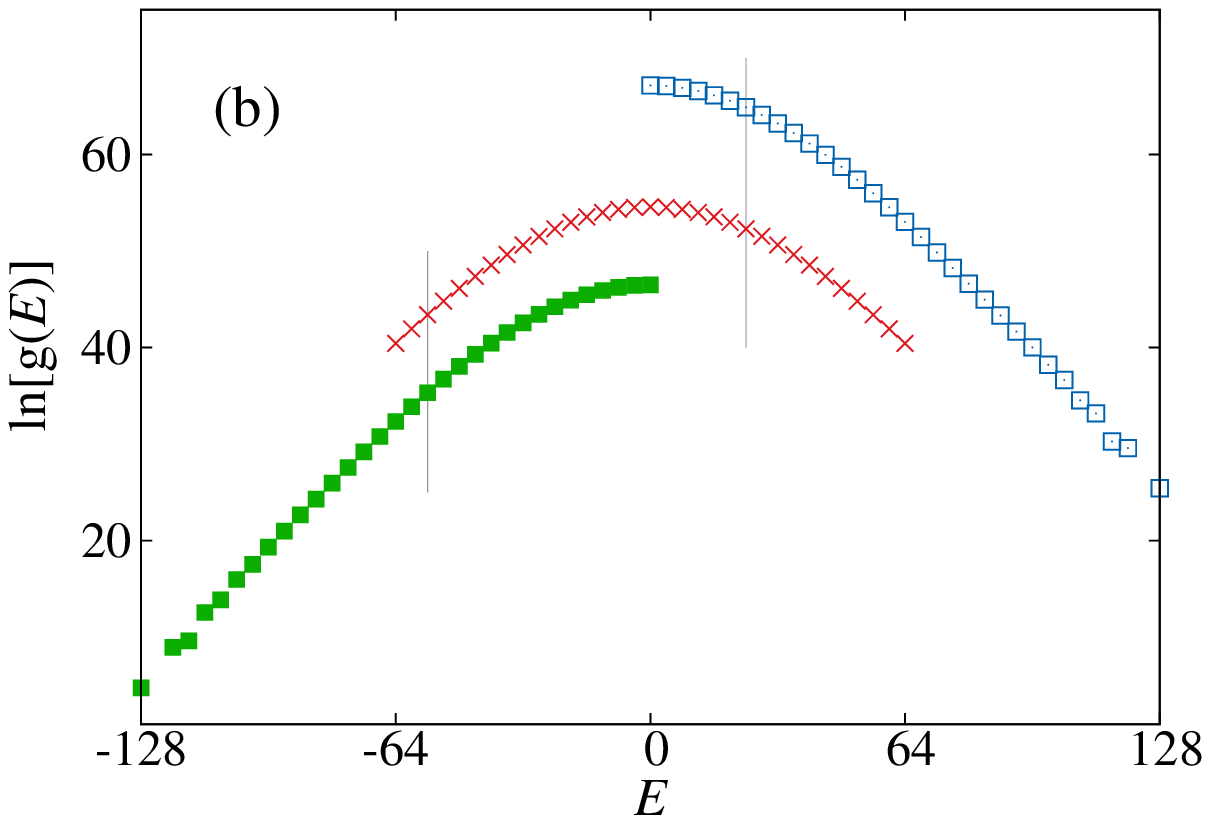}
\end{minipage} 
\caption{(color online) (a) Final $g(E,M)$ pieces of the two-dimensional Ising model on a $8\times8$ square lattice. DOS pieces are shifted upward progressively for visualization purpose. 
(b) Individual pieces of $g(E)$ obtained by a summation over the magnetization range (Eq. (\ref{sum_M})). Vertical lines indicate the matching points ($E_{\rm mat}$) for joining the DOS pieces together. $E_{\rm mat}$ is determined as the point with minimal difference in the inverse microcanonical temperatures of the two DOS pieces. See text for more details.
}
\label{fig2}
\end{figure}

At this stage of the simulation, each walker is left with a joint DOS fragment (Fig.\ref{fig2}(a)) and one needs to combine all the fragments to recover $g(E,M)$ over the entire phase space. To do so, we first sum over the magnetization and obtained the density of states as a function of energy:
\begin{equation}
g(E) = \sum_{M}{g(E,M)},
\label{sum_M}
\end{equation}
as shown in Fig.\ref{fig2}(b). After that we calculate the derivative of $\ln[g(E)]$ as a function of $E$ in the overlapping region. In this procedure, one finds a point $E_{\rm mat}$  where the difference between the inverse of the microcanonical temperature $\beta (E)= {\rm d} \ {\ln}[g(E)]/{\rm d}E $ is a minimum. This $E_{\rm mat}$ point is used to match and adjust the two adjacent surfaces. That is, the whole DOS surface $g(E,M)$ has to be rescaled at $g(E_{\rm mat},M)$ and the joint DOS is obtained by:
\begin{equation}
g(E,M)=\left\{\begin{array}{rc}
g_i(E,M),&\mbox{if}\quad E < E_{\rm mat},\\
g_j(E,M), &\mbox{if}\quad E \ge E_{\rm mat}.\end{array}\right.
\label{p-wl}
\end{equation}
where $i$ and $j$ are,  respectively, walkers in the lower and higher energy range and, $g_j(E,M)$ is the shifted value of $g(E,M)$ for the $j$th walker.

In principle, other ways of matching the 2D-DOS could be applied  \cite{REWLjop,REMUCAandMUCARE} and other ways to split the phase space might be desired to increase the performance of the simulation ( see, e.g., Ref. \cite{REWLprl} ). For example, one might be interested in splitting the magnetization instead of the energy range, or splitting both energy and magnetization ranges. In this case both dimensions of the overlapping phase space will need to be analyzed, one at a time, to find the $E_{\rm mat}$ and $M_{\rm mat}$ matching points.
Moreover, since multiple walkers can be used to sample a window simultaneously, it is possible to employ jackknifing or bootstrapping techniques to compute the statistical errors from a unique production run.

\section{Preliminary results for the two-dimensional Ising model}

From our simulations for the two-dimensional Ising model we examine ($i$) the generalization of the REWL approach to higher dimensions and ($ii$) the accuracy of the algorithm.
We carried out serial 2D-WL and 2D-REWL simulations by performing random walks in total energy $E$ and magnetization $M$ space until the modification factor reaches $f_{\rm final}=\textrm{exp}(10^{-8})$.  We adopted the $80\%$ flatness criterion, i. e., all entries in the histogram $H(E,M)$ are no less than the chosen percentage of the average height of the histogram.

\begin{figure}[h]
\centering
\begin{minipage}{19pc}
\includegraphics[width=19pc]{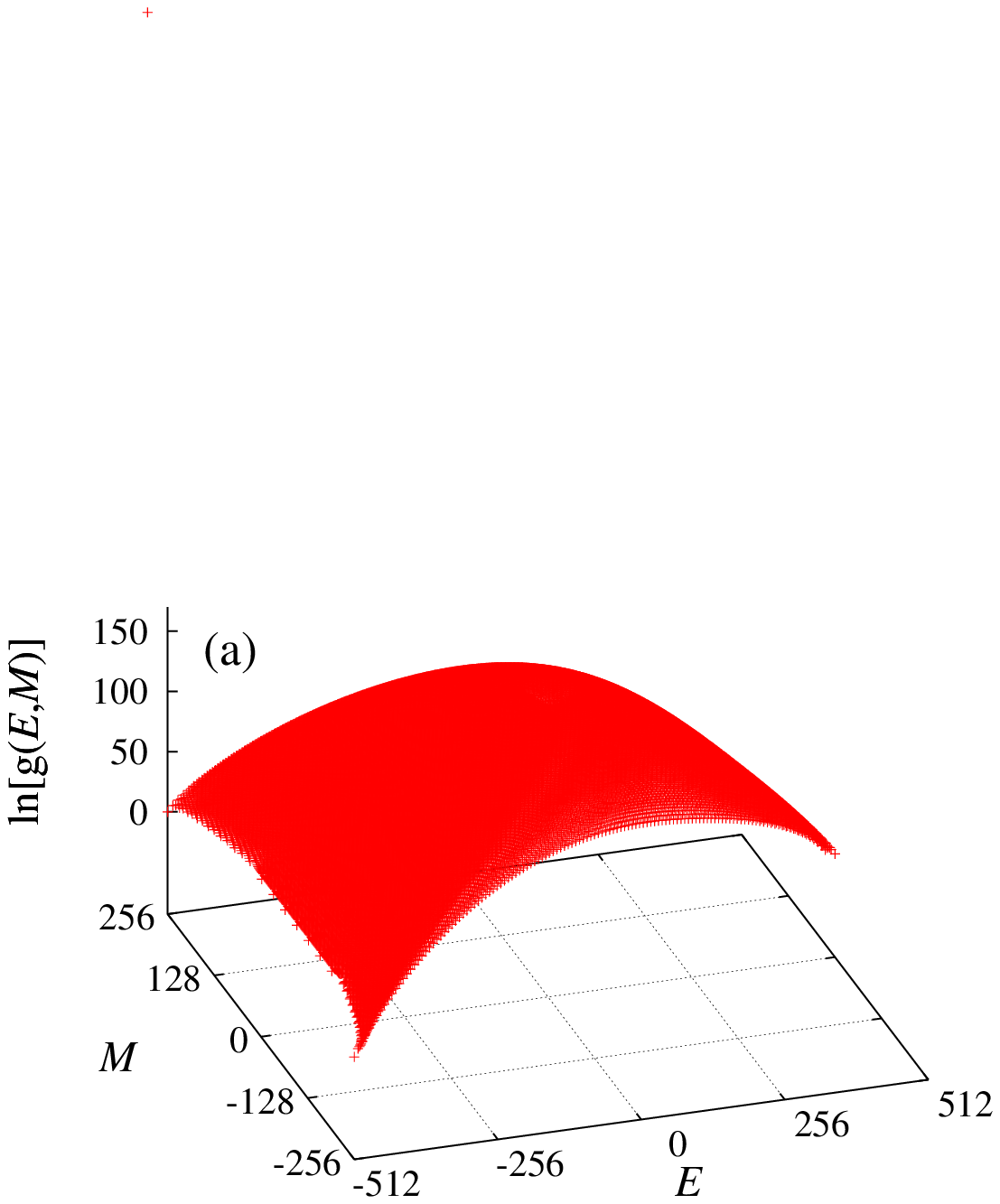}
\end{minipage}\hspace{0pc}%
\begin{minipage}{18pc}
\includegraphics[width=18pc]{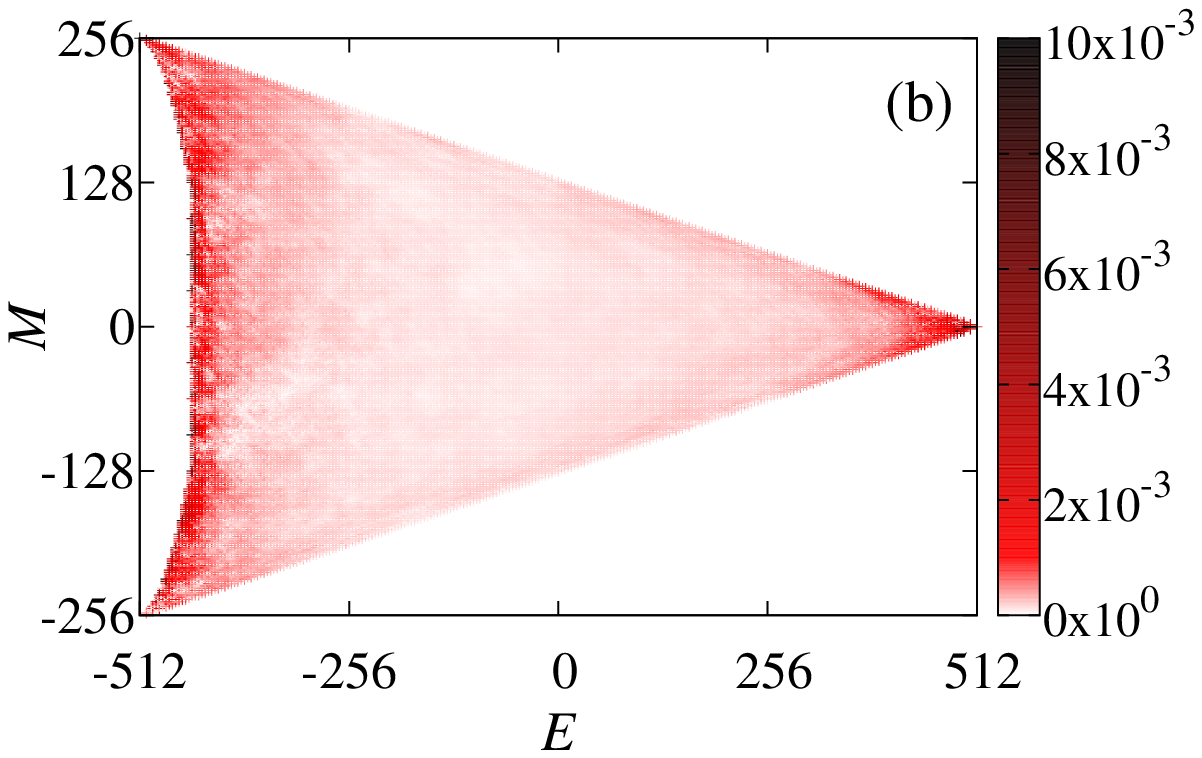}
\end{minipage}
\caption{(color online) Logarithm of the complete joint DOS for the Ising model on a $16\times 16$ square lattice obtained by the two-dimensional REWL approach. (b) Relative difference between the joint DOS obtained from WL and REWL sampling.
}
\label{fig4}
\end{figure}

For the 2D-REWL simulations we divided the parameter space into equal-sized windows with restricted values of $E$ and unrestricted $M$. We used a constant $50\%$ overlap between neighboring windows and multiple random walkers to keep track of the joint DOS within each window. Fig.\ref{fig4}(a) shows the resulting DOS that is obtained in a few hours by $9$ parallel walkers (the energy is split into $3$ windows with three walkers per window). An equivalent serial WL sampling over the entire two-dimensional parameter space with a single random walker takes several days to complete.

\begin{figure}[h]
\centering
\begin{minipage}{18pc}
\includegraphics[width=17pc]{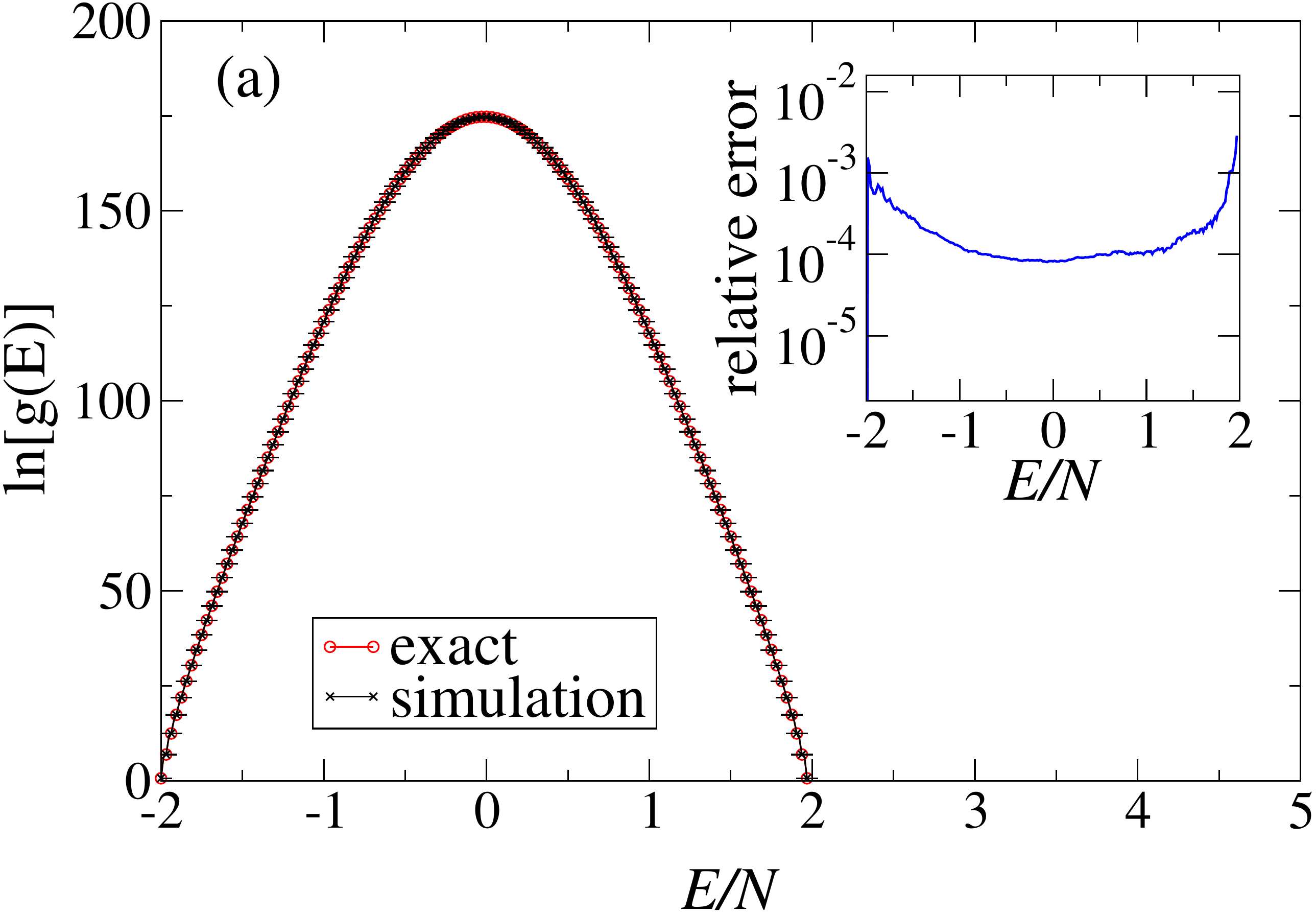}
\end{minipage}\hspace{1pc}%
\begin{minipage}{18pc}
\includegraphics[width=17pc]{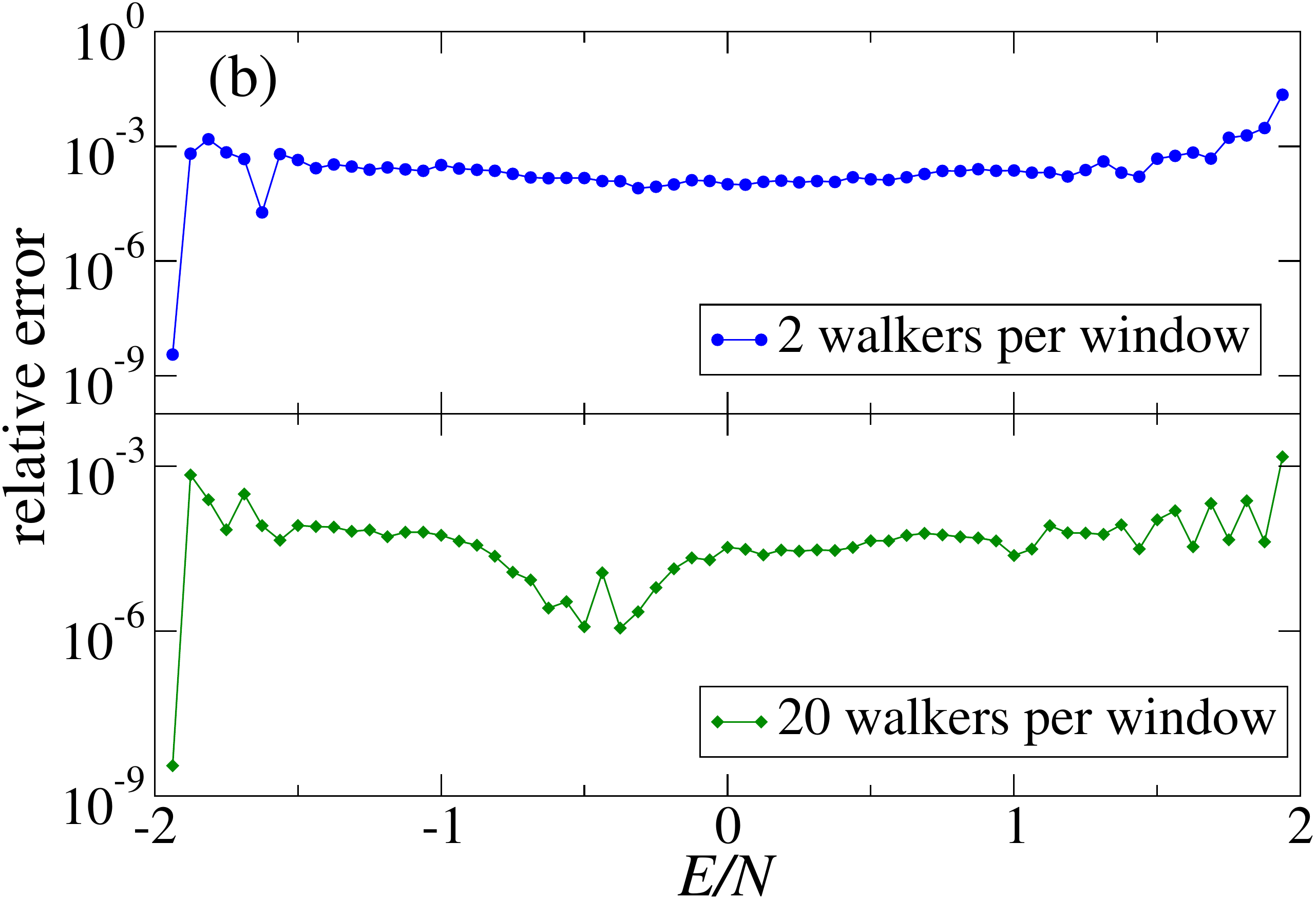}
\end{minipage} 
\caption{(color online) (a) Comparison of the density of states from 2D-REWL sampling and the exact results for a $16\times 16$ square lattice (relative errors are shown in the insert). (b) Relative errors for a $8\times 8$ lattice square using $2$ walkers per window (upper panel) and $20$ walkers per window (lower panel) respectively.}
\label{fig5}
\end{figure}

To compare the DOS obtained from WL and REWL sampling (denoted by $g_{\rm WL}(E,M)$ and $g_{\rm REWL}(E,M)$, respectively), we calculated the relative difference:
\begin{equation}
\Delta_{\rm diff}(E,M) = \frac{\left |g_{\rm WL}(E,M) - g_{\rm REWL}(E,M) \right|}{\bar{g}(E,M)},
\label{rel_diff}
\end{equation}

\noindent
where $\bar{g}(E,M) = (g_{\rm WL}(E,M) + g_{\rm REWL}(E,M)) / 2$. The relative difference is plotted in Fig. \ref{fig4}(b). Over the entire phase space, there is no apparent sign of systematic errors due to the matching of windows. The order of magnitude of the relative difference is comparable to the relative errors (as shown in Fig. \ref{fig5}(a)); they both share the same behavior of having larger values in the lower and upper energy edges. This is conceivable because of the rareness of configurations as well as the presence of cusps and empty regions in the phase space.

To verify the accuracy of the algorithm, we compute $g(E)$ with Eq. (\ref{sum_M}) so that we can compare with the exact results. Fig.\ref{fig5}(a) shows the logarithmic DOS obtained by using $3$ walkers per window in a 2D-REWL simulation along with the exact values, the relative errors are shown in the insert. We can see that the results obtained from our simulations agree extremely well with the analytic solutions. 
Fig.\ref{fig5}(b) shows the relative errors for the simulations employing $2$ walkers per window and $20$ walkers per window respectively. We observe that increasing the number of walkers, which update the DOS independently, decreases the sampling errors correspondingly. 

\hspace*{2em}

\section{Summary}
We implemented a two-dimensional replica-exchange Wang-Landau sampling method
and applied it to determine the joint density of states $g(E,M)$ for the 
Ising model on a square lattice. The joint DOS over the 
entire two-dimensional parameter space, obtained by combining the joint DOS
in the different sampled windows, agrees very well with the results obtained
with a serial WL sampling over the entire two-dimensional parameter space. No artifacts in
the final joint DOS were observed due to the parallelization and DOS joining procedures. 
We also showed that errors can be reduced by employing multiple random walkers in 
each window. Similar to the original REWL method, the higher-dimensional 
implementation of the REWL method is very efficient and it scales well with
the number of CPUs. Hence, it can be used to study much larger systems than
what is feasible using a serial WL method.

\section*{Acknowledgments}
This research was sponsored in part by the National Science Foundation under Grant OCI-0904685, by the Office of Science, Office of Basic Energy Sciences, Materials Sciences and Engineering Division (M.E.) and by the Office of Advanced Scientific Computing Research (Y.W.L.) of U.S. Department of Energy, under Contract DE-AC05-00OR22725. The simulations were carried out on resources provided by the Georgia Advanced Computing Resource Center at the University of Georgia. A. Valentim was sponsored by CAPES Foundation, Ministry of Education of Brazil, Brasilia - DF 70.040-020, Brazil.


\section*{References}
\begin{thebibliography}{9}
\bibitem{WLprl}
Wang F and Landau D P 2001 {\it Phys. Rev. Lett.}  {\bf 86} 2050

\bibitem{WLpre}
Wang F and Landau D P 2001  {\it Phys. Rev. E}. {\bf 64} 056101

\bibitem{WLajp}
Landau D P,  Tsai S -H and Exler M 2004 {\it Am. J. Phys.} {\bf 72} 1294

\bibitem{rathore} Rathore N and  de Pablo J J 2002 {\it J. Chem. Phys.}  {\bf 116} 7225

\bibitem{yamaguchi} Yamaguchi C and Kawashima N 2002 {\it Phys. Rev. E} {\bf 65} 056710

\bibitem{taylor} Taylor M P, Paul W and Binder K 2009 {\it  J. Chem. Phys.}  {\bf 131} 114907

\bibitem{wuest} W\"ust T and Landau D P 2009 {\it Phys. Rev. Lett.}  {\bf 102} 178101

\bibitem{REWLprl}
Vogel T, Li Y W, W\"ust T and Landau D P 2013 {\it Phys. Rev. Lett.}  {\bf 110}  210603 

\bibitem{REWLpre}
 Vogel T, Li Y W, W\"ust T and Landau D P 2014 {\it Phys. Rev. E} {\bf 90} 023302 

\bibitem{REWLjop} Li Y W, Vogel T, W\"ust T and Landau D P 2014 {\it J. Phys.:Conf. Ser.}  {\bf 510} 012012

\bibitem{TYVL} Vogel T,  Li Y W,  W\"ust T and Landau D P 2014 {\it J. Phys.:Conf. Ser.}  {\bf 487} 012001

\bibitem{marinari} Marinari E and  Parisi G 1992 {\it Europhys. Lett.} {\bf 19} 451

\bibitem{nemoto} Hukushima  K and  Nemoto K 1996 {\it J. Phys. Soc. Jpn.}  {\bf 65} 1604 

\bibitem{zhou} Zhou C,  Schulthess T C, Torbr\"ugge S and Landau D P  2006 {\it Phys. Rev. Lett.}  {\bf 96} 120201

\bibitem{asymising}
Tsai S -H, Wang F and Landau D P  2009 {\it Int. J. Mod. Phys. C} {\bf 20} 1357

\bibitem{pla} Silva C J,  Caparica A A and Plascak J A 2006 {\it Phys. Rev. E} {\bf 73} 036702

\bibitem{REMUCAandMUCARE} Sugita Y and  Okamoto Y 2000 {\it Chem. Phys. Lett.} {\bf 329} 261

\bibitem{beale} Beale P D 1996 {\it Phys. Rev. Lett.} {\bf 76} 78

\end{thebibliography}
\end{document}